\documentstyle[12pt]{article}

\def\be{\begin{equation}}
\def\ee{\end{equation}}
\def\ba{\begin{array}{c}}
\def\ea{\end{array}}

\def\ben{\[ }
\def\een{\] }

\begin{document}

\titlepage
%\vspace*{4cm}

\begin{center}

%\vspace{4.truecm}

{\Large \bf Asymptotic solvability of an imaginary cubic
oscillator with  spikes
 }

\vspace{1.truecm}

{\bf Miloslav Znojil}

Department of Theoretical Physics,

 Institute of Nuclear Physics,
Academy of Sciences,

 250 68
\v{R}e\v{z}, Czech Republic

e-mail: znojil@ujf.cas.cz

% \vspace{1.truecm}

 \vspace{0.5truecm}

%\vspace{2.truecm}

{\bf Franti\v{s}ek Gemperle}

 Department of Mathematics,
 Faculty of Nuclear Science and Physical Engineering,

 Czech Technical University

 Trojanova 13,
 120 00 Prague,
 Czech Republic

e-mail:  gefa@km1.fjfi.cvut.cz

% \vspace{1.truecm}

 \vspace{0.7truecm}

  and

 \vspace{0.5truecm}

 {\bf Omar Mustafa}

Department of Physics,
 Eastern Mediterranean University,

 G. Magosa, North Cyprus, Mersin 10 - Turkey

e-mail: omar.mustafa@mozart.emu.edu.tr

\end{center}

 % \vspace{0.5truecm}

%\today

\vspace{0.5truecm}

\section*{Abstract}

For complex potentials $V(x) = -(ix)^3 -\beta^2(ix)^{-2}
-2\,\beta\,\delta\, (ix)^{1/2} $ which are ${\cal PT}$ symmetric
we show that in the $\beta \gg 1$ strong coupling regime the low
lying bound states almost coincide with harmonic oscillators
whenever the spectrum remains real (this means, at all $\delta <
\delta_{critical}(\beta) \approx 1 $).

 \vspace{0.5truecm}

PACS 03.65.Fd   03.65.Ca  03.65.Ge 03.65.Bz

\newpage

\section{Introduction}

Radial Schr\"{o}dinger equations
 \begin{equation}
 \left[ -\,\frac{d^{2}}{dr^{2}}+ V_{(eff)}(r)
 \right ] \,\psi (r)=E\psi (r), \ \ \ \ \ \ \ V_{(eff)}(r) =
 \frac{\ell (\ell +1)}{r^{2}} + V(r)
 \label{schrod}
\end{equation}
characterized by the {\em large} angular momenta  $\ell$ appear in
molecular or nuclear physics~\cite{Mor}. The strongly repulsive
phenomenological core $G/r^{2}$ of $V(r)$ with $G \gg 1$ is often
added directly to the centrifugal term in three dimensions,
$\ell(\ell+1) = \tilde{\ell}( \tilde{\ell}+1) + G$, $\tilde{\ell}
= 0, 1, \ldots$ and, in the latter context, even the effective
dimensions $D\neq 3$ in $\tilde{\ell}=m+(D-3)/2$, $m = 0, 1,
\ldots$ may become extremely large~\cite{Sotona}.

Similar situations have inspired the development of a few
efficient $\ell \gg 1$ approximation techniques applicable to many
particular Hermitian models (cf. ref. \cite{Omma} containing many
further references). Thorough tests of their reliability are amply
available~\cite{Omar2}. Similar studies are missing in the
non-Hermitian context, and our present purpose is to fill the gap.

For the sake of definiteness, let us recollect the popular
anharmonic oscillator
\be
V_{eff}(r) = \frac{\ell (\ell +1)}{r^{2}}
 + \omega^2r^2 + g\,r^N, \label{effe}
  \label{radkv}
 \ee
which has been studied in many papers at large $\ell$. The
encouraging results of these studies (cf. \cite{Guardiola} giving
further references) show that the effective potential possesses a
deep minimum at certain {\em real} coordinate $r_0> 0$. Near this
value, the effective potential is very well approximated by the
{\em exactly solvable} harmonic oscillator. Thus, choosing
$\omega=0$ and $N=4$ for definiteness we have
 \ben V_{eff}(r) = V_{eff}(r_0) +
\frac{1}{2}\,V_{eff}''(r_0) {(r-r_0)^{2}} + {\cal O}(r-r_0)^3, \ \
\ \ \ \ r_0=\sqrt[6]{\frac{\ell(\ell+1)}{2g}}\
  \label{effendi}
 \een
where the higher-order contributions prove small.

The latter Hermitian example is a key guide to our present study.
We shall just ask what happens when the potential and/or
coordinates cease to be real in a way proposed, e.g., by Bender
and Boettcher~\cite{BB}. The deepest {physical} motivation of the
similar generalizations may probably be found in field theory
where the {\em non-Hermitian} oscillators emerge in the most
natural manner (cf. concluding remarks in \cite{BS} or recent
considerations in~\cite{BM}).

One of the first explicit studies of a non-Hermitian model
(\ref{radkv}) has been offered by Caliceti et al \cite{Caliceti}
who revealed the amazing reality of energies in the manifestly
complex potential $V$ with cubic anharmonicity $N=3$ and with a
purely imaginary coupling $g$ in one dimension~(i.e., still at the
real coordinates $r$ and with the parity $\ell=-1, 0$). For a long
time, this result has been treated as a mere isolated curiosity in
the literature, in spite of the well known existence of its fully
natural analytic continuation to the whole domain of analyticity
of the potential, i.e., into the cut complex plane of
coordinates~$r$. The details concerning the purpose of such a
complexification have been summarized, e.g., in
ref.~\cite{Alvarez}.

A quartic, $N=4$ parallel of the cubic model emerged a few years
later in connection with the puzzling coincidence (up to a sign)
between energy corrections in two {different} perturbative
models~\cite{Seznec}. An explanation has been given by Buslaev and
Grecchi \cite{BG} who discovered that the connection was mediated
by an auxiliary non-Hermitian potential (\ref{radkv}) with  $N=4$,
negative~$g$ and the most suitable complex choice of the
coordinate $ r = x - i\varepsilon$ where $x\in (-\infty,\infty)$.
Its purely real and discrete spectrum was proved to be bounded
below.

Both the above non-Hermitian anharmonic oscillator examples share
the commutativity of their Hamiltonian with the product of
operations ${\cal P}x{\cal P} = -x$ and ${\cal T} i {\cal T} = -i$
interpreted as parity and time reversal, respectively,
\be
H = {\cal PT}\,H\,{\cal PT} \equiv H^\ddagger\ .
 \label{ptsymmetry}
 \ee
The relevance of such a type of anti-linear symmetry became
appreciated only after Bender and Boettcher \cite{BB} re-analyzed
the ${\cal PT}$ symmetric potentials (\ref{radkv}) with complex
couplings $g \sim i^N$. Putting $\omega=0$ for simplicity they
found that the numerical and WKB analysis supports the hypothesis
that the spectrum remains discrete, bounded below and purely real
at {\em any} real exponent $N \geq 2$. This inspired an intensive
further research and, within the resulting conjecture of the so
called ${\cal PT}$ symmetric quantum mechanics~\cite{BBjmp}, the
validity of eq. (\ref{ptsymmetry}) has been interpreted as a
certain analogue or a weaker form of Hermiticity
\cite{BG}-\cite{WH}.

In what follows we constrain our attention to the $N=3$
anharmonicity and pick up a one-parametric ${\cal PT}$ symmetric
generalization of eq. (\ref{radkv}) in section \ref{II} and
subsection \ref{IIa}. In subsection \ref{cunum} we then recollect
a few basic formal ingredients and, in particular, explain the
relation between the ${\cal PT}$ symmetry and boundary conditions.
A way in which the smallness of $1/\ell$ may play a crucial role
is outlined in the subsequent section \ref{num}. In section
\ref{tri} we describe our main results concerning the closed
asymptotic representation of the energies in the two separate
(viz., weak- and strong-coupling) regimes of the non-Hermitian
cubic-plus-square-root models. A few numerical tests and
illustrations confirm our assertions in section \ref{IV} and are
complemented by a non-numerical discussion in section \ref{V}. A
short summary follows in section~\ref{VI}.

\section{The problem \label{II}}

\subsection{Characteristic example: Cubic oscillator with spikes
 \label{IIa}}

The short history of the  ${\cal PT}$ symmetric quantum mechanics
climaxed with the recent work by Dorey, Dunning and Tateo (DDT,
\cite{DDT}) who succeeded in rigorously {proving} that the
generalized cubic model with $\omega=0$, viz.,
\begin{equation}
\left[ -\,\frac{d^{2}}{dr^{2}}+\frac{\ell (\ell +1)}{r^{2}}-\alpha
\,\sqrt{ i\,r}+i\,r^{3}\right] \,\psi (r)=E\psi (r), \ \ \ \
 \ r = x - i\varepsilon, \ \ \ x \in (-\infty,\infty)
 \label{rad}
\end{equation}
possesses the real and discrete spectrum whenever the
angular momentum $\ell$ is sufficiently large,
\begin{equation}
\ell\  >  \ \max \left[ \frac{1}{4}(2\alpha
-7),-\frac{1}{2}\right] \ . \label{bpou}
\end{equation}
The reality of the spectrum under a suitable constraint is a
characteristic feature of many pseudo-Hermitian
models~\cite{Mostafazadeh}. Within this class, the ${\cal PT}$
symmetric DDT oscillator (\ref{rad}) is distinguished by the
presence of two spikes near the origin. Moreover, the preliminary
tests performed on a simplified model in ref. \cite{DDT} indicate
that the reality of the spectrum becomes spontaneously broken. At
some $\ell_{minimal}(\alpha )$, only slightly below the bound
(\ref{bpou}), at least two energy levels have to merge and form a
complex conjugate pair. In the other words, we may tolerate the
non-Hermiticity of the Hamiltonian $H^{(DDT)}$ as {\em acceptably
weak} in the semi-infinite interval of the angular momenta $\ell
\in (\ell _{minimal},\infty )$ where $\ell _{minimal}$ grows with
$\alpha$ for $ \alpha>5/2$.

The interference between the two spikes becomes particularly
interesting in the strong-coupling domain of $\alpha \gg 1$. In
this interval, the square-root dynamics may be understood as
weakly non-Hermitian (in the sense of generating the real
spectrum) if and only if the kinematical centrifugal repulsion
remains also strong, $\ell = {\cal O}(\alpha)$. The latter
observation has attracted our attention since the non-Hermiticity
may in general worsen the feasibility of the construction of the
solutions while, as we already noted, many of the difficulties
with the construction of the bound states may in principle be
avoided due to the presence of the small parameter~$1/\ell$.

\subsection{Complex boundary conditions  \label{cunum}}

The $\ell-$dependence in the Schr\"{o}dinger equations remains the
same for the real and complex potentials. The latter case is
exemplified by our eq. (\ref{rad}) and illustrated in Figure~1
where, for simplicity, the smooth term $i r^3$ is completely
omitted.  The real part of the effective potential is displayed
there for the complex shift $\varepsilon = 0.8$ and angular
momentum $\ell=13/4 \approx \ell_{minimal}(\alpha)$ at the medium
$\alpha = 10$.  The Figure shows how the square-root force
$-\alpha\,\sqrt{ i\,r(x)}$ dominates at the large coordinates
while the centrifugal spike prevails near the origin.  Both these
spikes would become more pronounced closer to the real axis and
{vice versa}.

With the wave functions unconstrained by boundary conditions we
may construct two independent solutions $\psi_{1,2}(r)$ of our
imaginary cubic Schr\"{o}dinger differential equation (\ref{rad})
which are analytic functions of $r$ and/or $x$. In a way described
in more detail, say, in ref. \cite{Alvarez} we may cut the complex
plane of $r=r(x)$ from the origin upwards. This means that we
parametrize
 \[
r=\xi\,\exp i\,\varphi, \ \ \ \ \ \  \xi \in (0,\infty ), \ \ \ \
\ \varphi \in (-3\pi /2,\pi /2).
 \]
Using such a polar representation of $r(x)$ we may assign the
unique meaning to the square root expression $\sqrt{i\,r(x)}$. In
accordance with Figure~1 the real part of this long-range spike is
oriented upwards, ${\rm Re}\,\sqrt{i\,r}\geq 0$.  This convention
makes our potential uniquely defined.  Its complex and
$\varepsilon-$dependent effective form reads
 \be
V_{(eff)}[r(x)] =\frac{\ell (\ell +1)}{r^{2}(x)}-\alpha
\,\sqrt{i\,r(x)}+i\,r^{3}(x), \ \ \ \ \ \ r(x)+i\,\varepsilon= x
\in (-\infty,\infty).
 \label{reads}
 \ee
Due to the analyticity of this function in the cut complex plane
we may freely deform the integration path. The spectrum
$E^{(DDT)}$ remains unchanged for all the constant shifts
$\varepsilon >0$ so that one may work with the Buslaev's and
Grecchi's \cite{BG} asymptotic boundary conditions $\psi
(-i\,\varepsilon \pm \infty )=0$. Bender and Boettcher \cite{BB}
emphasized that a further generalization of these boundary
conditions may be admitted and reads
\[
\psi (\xi e^{i\,\varphi _{+}})=\psi (\xi e^{i\,\varphi _{-}})=0,\
\ \ \ \ \ \xi \rightarrow +\infty,
\]
\be
\varphi_{+}\in (-3\pi /10,+\pi /10),\ \ \ \ \ \ \varphi_{-}\in
(-11\pi /10,-7\pi /10). \label{wedge}
\ee
All of these boundary conditions are mutually equivalent and
form an elementary analytic continuation of their standard
special case $\psi (\pm \infty )=0$ in the wedge-shaped vicinity
(\ref{wedge}) of both the ends of the real axis. This is
illustrated in Figure~2 where the asymptotic wedge permitted for
the cubic oscillator is marked by the symbol CO. Its boundary
also avoids the upwards-running cut which starts at~$r=0$.

In the context of the textbook quantum mechanics, the removal of
the origin $r=0$ from our considerations has the two immediate
consequences. Firstly, in the spirit of ref. \cite{BG} (where more
details may be found) we may always return to the current
Hermitian radial-equation case (in more dimensions, with $r \in
(0, \infty)$) by a suitable limiting transition ($\varepsilon \to
0$ in our above notation). In this sense, one need not change the
mathematical results but discards merely ``a half" of the
available solutions as ``manifestly unphysical" due to the
divergence of their norm in this limit. Thus, for example, one
simply crosses out all the quasi-even states in the solvable
example of ref.~\cite{ptho}.

Secondly, there exists an alternative physical context where the
presence of the spikes of the form  $\ell(\ell+1)/r^2$ is a
dynamical assumption~\cite{Tater}. Then, the Hermitian quantum
system usually stays to be defined on the whole real axis, $r \in
(-\infty,\infty)$. All the solutions retain their physical meaning
even after the limiting transition $\varepsilon \to 0$ which
merely represents a regularization recipe. Such a regularization
is also needed within the so called supersymmetric quantum
mechanics~\cite{last}.

\section{The method \label{num}}

As we have seen, the non-Hermiticity is most easily introduced in
a Schr\"{o}dinger equation by a downward complex shift of the
coordinate. The typical consequences of this complexification may
be illustrated via the simplest example (\ref{radkv}) with the
vanishing $g=0$ and, say, scaled-out $\omega=1$. This offers us a
suitable guide towards the $1/\ell$ approximations in their
non-Hermitian generalizations.

Although the ${\cal PT}$ symmetric  $g=0$ oscillator is exactly
solvable and its spectrum is real, it still exhibits certain
unusual features~\cite{ptho}. The energies remain non-equidistant
and have to be numbered by the integer $n=0, 1, \ldots$ {\em and}
by the superscript~$^{(\pm)}$,
\be
E_n^{(\pm)}=4n+2\pm (2\ell+1). \label{exho}
 \ee
Once we abbreviate $ [\ell(\ell+1)]^{1/4}=A=A(\ell)> 0$ and assume
that this quantity is large, we arrive at the Schr\"{o}dinger
equation
\begin{equation}
\left[ -\,\frac{d^{2}}{dr^{2}}+  \frac{A^4}{r^{2}} +r^2
 \right] \,\psi (r)= E\,\psi (r).
 \label{adkv}
\end{equation}
We may infer that the absolute minimum of
$V_{(eff)}(r)={A^4}/{r^{2}} +r^2$ lies at the purely imaginary
point $r = R_0= -i\,A$. In the domain of our interest with the
large values of $A \gg 1$, it may be re-written in a perturbative
form using the new, shifted variable $t = r - R_0$,
\begin{equation}
\left[ -\,\frac{d^{2}}{dt^{2}} - 2\,A^2+4\,t^2+ {\cal O} \left(
\frac{t^3}{A} \right )
 \right] \,\psi  \left(t+R_0 \right )=E\,\psi  \left(t+R_0 \right ).
 \label{dkv}
\end{equation}
Obviously, the textbook perturbation solution of this problem is
straightforward \cite{Fluegge} and shows that the contribution of
the corrections is asymptotically small. We get the following
harmonic-oscillator leading-order energy estimate for the
low-lying part of the spectrum,
 \be
 E=-2\,A^2+2\,(2n+1)+{\cal O} \left(
\frac{n^2}{A^2}
 \right )
 , \ \ \ \ \ n = 0, 1, \ldots\ .
 \label{nase}
 \ee
It is worth emphasizing that our {\em non-Hermitian} equation
degenerates to its {\em Hermitian} harmonic oscillator
approximation which does not contain any centrifugal barrier.

The numerical reliability of eq. (\ref{nase}) is documented by the
third column in Table~1. In the light of the Table the harmonic
oscillator approximation reproduces the low-lying part of the toy
spectrum $E_n^{(-)}$ with reasonable quality even at a very small
$A = \sqrt[4]{30} \approx 2.34$.

Of course, the second, quasi-odd series $E_n^{(+)}= 2\ell
+4n^{(+)}+3 $ of energies is not reproduced here at all. The
reason is given by the error term in eq. (\ref{nase}). As soon as
we moved in the domain of the extremely large $\ell \to \infty$,
the value of the very lowest quasi-even excitation energy
$E_0^{(+)}= 2\ell+3$ is already comparable with our error
estimate. The high-lying energy $E_0^{(+)}$ cannot be reproduced
within the framework of our harmonic oscillator fit.

\section{Energies $E^{(DDT)}$ at the large $\ell$ \label{tri}}

The main merit of the freedom in the choice of the shift
$\varepsilon>0$ in our non-Hermitian Schr\"{o}dinger equation as
well as in the related boundary conditions is that we may let the
axis $r(x)$ pass through a minimum of the {\em complex}
interaction term. A full analogy with the Hermitian case is
achieved in this manner. The systematic search for all the
possible {\em complex} extremes of the effective potentials is
easy and may be based on the elementary mathematical rule
\begin{equation}
 \partial _{r}V_{(eff)}(r)|_{r=R}=0
. \label{oposi}
\end{equation}
In the vicinity of the extreme we may approximate the unsolvable
potentials by their reduction to the solvable harmonic oscillator
wells.

\subsection{Weak-coupling domain, $\alpha \ll \ell$
 \label{weak}}

As long as the value of $\ell $ is assumed to be very large in our
particular DDT example, it makes sense to abbreviate $2\ell (\ell
+1)/3=L^{5}$ and replace $\ell$ by the alternative large parameter
$L=L(\ell) \gg 1$. The range of $\alpha$ is then limited by the
condition (\ref{bpou}) so that we may re-scale $\ \alpha
=\sqrt{6L^5}\ \delta \ $ with $0 \leq \delta \stackrel{\textstyle
< }{\sim} 1$. This simplifies our Schr\"{o}dinger equation
(\ref{rad}),
\begin{equation}
 \left[ -\,\frac{d^{2}}{dy^{2}}+L^5\,
 W(y) \right ]\,\psi (r)=L^2 E\psi (r), \ \ \ \ \ \ \ r = L \,y\ .
 \label{radda}
\end{equation}
The exact form of the re-scaled effective potential
\[
 W(y)=
 \frac{3}{2y^{2}}-\delta \,\sqrt{6\, i\,y}+i\,y^{3}
\]
is $L-$independent. This simplifies the implicit definition of the
minimum/minima of $V_{(eff)}(r)$ which re-scales with ${r}=L\,y $
to an elementary formula
\begin{equation}
 \partial _{y}W(y)|_{y=Y}=0
\label{fareeaday}
 \end{equation}
In writing the solutions of this equation we shall distinguish
between the two separate intervals of~$\alpha$. In the
weak-coupling regime with the negligible $|\alpha| \ll \ell \ $
(i.e., vanishing $\delta \approx 0$), equation (\ref{fareeaday})
is trivial ($Y^5=-i$) and implies that the extremes of $
V_{(eff)}(r) $ are located at the five complex points numbered by
$k=0,1,2,3,4$,
\[
{r}={R}_{k}=\left| \left[ \frac{2}{3}\ell (\ell
+1)\right]^{1/5}\right| \,Y_{k},\ \ \ \ \
Y_{k}=-i\,e^{i(-8+4k)\pi /10},\
\ \ \
\delta =0.
\]
After the overall change of scale of $ {r}=L\,y\ $ with $L={\cal
O}(\ell^{2/5} )\gg 1$ we get our effective potential
$W(y)=V_{(eff)}(L\,y)/L^3$ re-written in the form of the Taylor
series,
\begin{equation}
W(y)=\frac{3/2}{y^{2}}+i\,y^{3}=\frac{5}{2Y_k^{2}}
 +\frac{15}{2Y_k^{4}}\,{\
(y-Y_k)^{2}}+{\cal O}\left[ (y-Y_k)^{3}\right] .
  \label{appr}
\end{equation}
As long as ${\cal R}e\ Y_k^{4}$ is negative if and only if $k = 0$
or $k=4$ while ${\cal I}m\ Y_k^{4}$ is non-zero unless $k=2$, the
{\em unique absolute minimum} of the effective potential $W(y)$
exists and lies at the point $y=Y_{2}=-i$. This implies that in
the first two orders in our auxiliary small parameter
$1/\ell^{const}\ll 1$ we get the low-lying spectrum
 \be
E_n^{(DDT)} \approx
-\frac{5\,L^3}{2}+\sqrt{\frac{15\,L}{2}}\,(2n+1), \ \ \ \ n = 0,
1, \ldots\ \label{enees}
 \ee
We see that the choice of the integration path $r(x)$ with
$\varepsilon = L= i\,R_2$ enables us to replace our differential
Schr\"{o}dinger eq. (\ref{rad}) by its harmonic oscillator
approximation. The asymptotic compatibility of the related
boundary conditions is illustrated by Figure~2.

\subsection{Strong-coupling regime, $\alpha ={\cal O}(\ell)$
  \label{weaker}}

The strength $\alpha$ of the long-range spike is constrained by
the rule (\ref{bpou}). This means that the ${\cal PT}$ symmetry
breaks down at the couplings $\delta_{critical} \approx 1$ in a
way described in ref. \cite{DDT}. The choice of $\delta\in
(0,\delta_{critical})$ remains compatible with the reality of the
spectrum and characterizes the strongly spiked regime where both
our free parameters are comparably large, $\alpha ={\cal
O}(\ell)$.

For the small non-vanishing $\delta\approx 0$ we may expect that
the positions $Y_k$ of the five weak-coupling extremes of $W(y)$
become only slightly shifted. In the vicinity of the absolute
minimum with $k=2$ we abbreviate $r=-iLq$ and write
\[
W(y) =3/ (2y^{2}) +iy^{3} -\delta\,\sqrt{6iy}= -\frac{3}{2\,q^2}
-\delta \sqrt{6\,q}-q^3.
\]
Equation (\ref{fareeaday}) remains quadratic in the fifth power of
the re-scaling factor in $R=-iLQ$,
\[
1-
\left (2+
\frac{\delta^2}{6}
\right )\,Q^5 + Q^{10}=0.
\]
We abbreviate $Q^5=Z$ and arrive at the two eligible roots. Both
of them are real and share the obligatory weak-coupling limit
$Z_{\pm }=1$. When $\delta$ grows from 0 to 1, they split and move
to their respective strong-coupling extremes,
\[
\left[ Z_{\pm }=1\right]_{^{_{_{\delta =0}}}}\ \ \longrightarrow \
\ \left[ Z_{+}=\frac{3}{2},\ Z_{-}=\frac{2}{3}\right]_{_{\delta
=1}}.
\]
In the light of our original eq. (\ref{fareeaday}) which may be
re-written in the form
\[
1-Z=
\delta \,\sqrt{\frac{Z}{6}},
\]
the larger root gives the wrong sign on the right-hand side and
must be discarded for positive $\delta$.  The solution becomes
unique and its value $Z=Z_{-}\leq 1$ decreases with the growing
$\delta$. Near its absolute minimum, our effective potential $W(y)
$ depends on $\delta$ only via the function $Z=Z(\delta)$,
\begin{equation}
W(y)=\frac{1}{Z^{2/5}} \left (-\frac{15}{2}+5~Z\right )
 +\frac{15}{4\,Z^{4/5}} \left (1+Z\right )\left (y+i\,Z^{1/5}
 \right )^2+\ldots\ .
 \label{jump}
\end{equation}
As long as $Z^{1/5}(1)\approx 0.922$, this is a fairly weak
dependence. The formula leads to our final energy estimate
\begin{equation}
E_{m}^{(DDT)}=\frac{1}{Z^{2/5}}\left[ \left (-\frac{15}{2}+5~Z
\right ) \,L^3+\sqrt{\frac{15~(1+Z)}{4}}\,L^{1/2}\,(2m+1) +\ldots
\right] \label{DDTE}
\end{equation}
with $ m=0,1,\ldots $. Near $\delta=0$ the deformation of the
spectrum (\ref{enees}) is continuous and smooth.

\section{Numerical tests \label{IV}}
%===================================

Let us support the idea of applicability of our formulae  by their
immediate comparison with exact results generated by a suitable
''brute-force" numerical method. All the necessary calculations
will be performed with the help of the discrete variable
representation (DVR) approach of Harris, Engerholm and
Gwinn~\cite{Harris65}. Since its use is rather new in the present
context, let us start from its brief description.

In the simplest case, the DVR approach can be regarded as a
variational method for finding bounded solutions of the
Schr\"odinger equation. As a first step, we have to choose a
suitable set of orthonormal functions $\{\phi_k(r)\}_{k=0}^N$ that
should be real for a $\cal PT$ symmetric case. Then we evaluate
matrix elements of the Hamiltonian:
  \begin{eqnarray}
  H_{ij} &=& T_{ij} + V_{ij} \ ,\\
  \label{tij}
  T_{ij} &=& -\int \phi_i(x) {d^2\over dx^2} \phi_j(x) \, dx \ ,\\
  \label{vij}
  V_{ij} &=& \int \phi_i(x) V(x) \phi_j(x) \, dx \ .
  \end{eqnarray}
The approximate bounded solutions and their energies are finally
obtained as eigenvectors and eigenvalues of the Hamiltonian
matrix~$H$.

Since the basis set is usually formed by special functions (e.g.
orthogonal polynomials), the former integrals~(\ref{tij}) can be
evaluated analytically. However, the evaluation of the latter
ones~(eq.~(\ref{vij})) is tedious. Therefore in the DVR, the
Hamiltonian matrix is expressed and approximated in another
representation. We calculate matrix elements of the coordinate
operator
  \begin{equation}
  X_{ij} = \int \phi_i(x) x \phi_j(x) \, dx
  \end{equation}
which mostly can be done analytically. Then we transform the
obtained matrix into its diagonal representation $\Lambda$:
  \begin{equation}
  X \to \Lambda = Q^{-1} X Q \ .
  \end{equation}
The Hamiltonian matrix is now approximated in the
$\Lambda$-representation as
  \begin{eqnarray}
  H(\Lambda) &=& T(\Lambda) + V(\Lambda) \ , \\
  T(\Lambda) &=& Q^{-1} T Q \ , \\
  V_{ij}(\Lambda) &\approx& \delta_{ij} V(\Lambda_{ii}) \ .
  \end{eqnarray}
The approximation basically arises from the fact that an
eigenvector of the coordinate matrix $X$ should correspond to a
function that is localized around the appropriate eigenvalue
$\Lambda_{ii}$. The level of the approximation is further
discussed in ref.~\cite{Dickinson68}.

The DVR can also be regarded as a method that allows to construct
wave functions in a set of discrete points, i.e. on a grid. For
each set of basis functions we obtain a corresponding grid, and an
appropriate approximation of the kinetic energy operator
$T(\Lambda)$. The grid can be shifted in the complex plane or even
scaled or rotated, provided that a corresponding scaling or
rotation is also performed with the kinetic energy operator. The
choice of basis functions affects only the spacing between
successive points. For example, with $\phi_k(x)\sim\sin kx$, the
grid is nearly equidistant while with the Hermite polynomials we
obtain a grid that is symmetrical and denser in the middle.

In our calculations we used Hermite polynomials as basis functions
(see e.g. ref. \cite{Horacek96} for a precise definition of this
DVR). The grid consisted from up to 1001 points from the interval
$x\in\langle-17.5,17.5\rangle$ with
$\epsilon\in\langle0.5,5\rangle$. The Hamiltonian matrix was
diagonalized with the help of the QR routine for general complex
matrices from the EISPACK package~\cite{eispack}. The results were
verified to be stable with respect to the changes of the grid.

As a case study let us first contemplate the exactly solvable eq.
(\ref{radkv}) at $g=0$. The sample of the exact energies is given
in the first column of Table~1. Their purely numerical DVR
reproduction appears in the second column and we see that the
method is fully reliable.

The last two columns of the Table illustrate the efficiency and
practical value of our non-numerical estimate (\ref{nase}). We
have to emphasize that even in the domain of the not too large
$\ell$, the difference between the exact and approximate energies
remains small, of the order of $\approx 0.5 \%$ for ground state.
The error term remains the same also for all the excited states
but this should be understood as a mere peculiarity of our choice
of the oversimplified illustrative example.

The genuine test of the present formulae (and of their merits as
well as limitations of validity) only appears in Table~2. The
''measure of smallness" $1/L$ lies there, roughly, in between
$0.55$ and $0.23$. This choice makes the test quite stringent
again. Its results are very encouraging. We witness, firstly, a
quick improvement of the quality of the ground state, from cca
$1\%$ at $L\approx 2$ up to the four-digit precision at $L \approx
4$. Secondly, the poor performance of our harmonic oscillator
approximation of excited states at $L\approx 2$ (giving an almost
$100\%$ error already for the second excited state) is in a sharp
contrast with the  $L \approx 4$ results predicting the reasonably
good two correct digits even for the $8$th excited state.

In Table 3 we extend the scope of our test beyond the
weak-coupling regime with $|\alpha| \ll 2\,\ell$ and/or
$\delta=0$. The dependence of the energies on the non-vanishing
values of the parameter $\alpha=2\sqrt{\ell(\ell+1)}\delta \approx
20.98\,\delta$ is shown there to agree very well with our
asymptotic prediction (\ref{DDTE}) even at the fairly small $L
\approx 2.36$. We may note that the precision of our approximation
appears to be almost $\alpha-$ independent in a broad range of
$\alpha$ including also the domain of the negative values which
were not discussed here in detail as safely protected against any
possible spontaneous ${\cal PT}$ symmetry breaking~\cite{DDT}.

\section{Discussion \label{V}}

At the maximal $\delta \approx 1$, the estimate of the low lying
energies
\begin{equation}
E_{m}^{(DDT)}\approx \left ( {\frac{3}{2}} \right )^{2/5}\,
 \left [
 -\frac{25}{6}\,L^3 + \frac{5}{2}\,
 \sqrt{L}\,(2m+1)
 \right ], \ \ \ \ \ m = 0, 1, \ldots\ , \ \ \ \ \ \
\delta \approx 1\ .
\end{equation}
differs significantly from its weak-coupling counterpart
(\ref{appr}). Moreover, in the light of the numerical experiments
of ref.~\cite{DDT} we may expect the end of the applicability of
our straightforward harmonic-oscillator approximation. At the same
time, the underlying shift of the minimum looks insignificant. We
may conclude that the freedom in our choice of the shift
$\varepsilon$ may encounter its natural limitations near and
beyond the value of~$\delta = 1$.

The latter point may comparatively easily be discussed
quantitatively. Returning to the re-scaled form of our original
Schr\"{o}dinger eq. (\ref{radda}) we may re-write this complex
differential equation in an equivalent form on real line,
\[
 \left\{ -\,\frac{d^{2}}{dx^{2}}+L^3\,
 \left [
 \frac{3}{2({x}/L-i\eta)^{2}}-\delta \,\sqrt{6\,
  i\,({x}/L-i\eta)}+i\,({x}/L-i\eta)^{3} \right ]
 \right \}\,\phi (x)= E\phi (x)\ .
\]
Let us assume now that the shift $\eta=\varepsilon/L$ is not too
small. This enables us to transform the effective potential in a
series in the powers of $L$,
\[
 \left\{ -\,\frac{d^{2}}{dx^{2}}+{L}^{3}
 \left ( - \frac{3}{2\,
\eta ^{2}}-\delta\, \eta ^{3}-{{\it \eta }} ^{3}\right ) +i
x{L}^{2}\left (
 \frac {3}{\eta ^{3}}
 -\frac{\delta\,\eta^{2}}{2}
 -3\, \eta^{2} \right )+ \right .
 \]
 \be
\left .
 +{x}^{2}L\left (
 \frac{9}{2\,\eta^{4}}-\frac {\delta\, \eta }{8}
 +3\,{\it \eta }\right ) + R
 \right \}\,\phi (x)= E\phi (x)\ . \label{eep}
 \label{radnet}
 \ee
The residual term can only generate ${\cal O}(1)$ corrections to
the energies and may be omitted as irrelevant. As a consequence,
we may once more shift the coordinate line, $x=z+i\,\varrho$ and
eliminate the redundant linear term in $z$ via the suitable choice
of $\varrho$. The new version of the potential reads
\be
{\frac {\left (36-\delta\,{ {{ \eta }}}^{5}+24\,{ {{ \eta
}}}^{5}\right )L}{{ {{ 8\,\eta }}}^{4}}}\,z^2 -{\frac {\left
(72+81\,\delta\,{ {{ \eta }}}^{5}+ 216\,{ {{ \eta
}}}^{5}-3\,{\delta}^{2}{ {{ \eta }}}^{ 10}+34\,\delta\,{ {{ \eta
}}}^{10}+12\,{ {{ \eta }}}^{ 10}\right )\,{L}^{3}}{{ {{2\, \eta
}}}^{2}\left (36-\delta\,{ {{ \eta }}}^{5}+24\,{{{ \eta
}}}^{5}\right ) }}. \label{see}
 \ee
The derivation of the subsequent $\eta-$dependent energies will be
skipped here as straightforward. Their role is less important in
the present context but may be expected to grow in the
perturbative context (of course, the explicit study of the
higher-order perturbation corrections lies already beyond the
scope of our present paper).

Formula (\ref{see}) depends on a free parameter $\eta$ and offers
a more flexible harmonic oscillator fit of the effective potential
in eq. (\ref{radnet}). Its $\eta-$dependence may certainly prove
useful in numerically oriented considerations, as it still gives
the approximate energies up to a bounded ${\cal O}(R)={\cal O}(1)$
error term. Non-numerically, formula (\ref{see}) may be checked as
reproducing {\em exactly} our previous eq. (\ref{appr}) in the
weak coupling limit $\delta = 0$ with optimal~$\eta = 1$.

A small decrease of $\eta = 1-\lambda$ may be contemplated as a
small perturbation. The deep, ${\cal O}({L}^3)$ minimum of our
$\eta-$dependent effective potential does not move in the first
order at all. Only the shape becomes narrower pushing the ${\cal
O}(\sqrt{L})$ component of the energies (\ref{enees}) upwards by
the factor $1/\eta \approx 1 + \lambda $.  This indicates a
certain variational optimality of our previous leading-order
results where the shift was in fact {\em not} arbitrary,~$\eta =
1$.

\section{Summary \label{VI}}

In this paper we paid attention to the complex and asymptotically
cubic DDT oscillator (\ref{rad}).  Within the so called ${\cal
PT}$ symmetric quantum mechanics, this oscillator represents one
of the most characteristic examples of a non-Hermitian (or ''next
to Hermitian") Hamiltonian with real spectrum. Although the model
is not solvable in closed form, its {appeal} is enhanced by the
presence of the variable coupling $\alpha$ and angular momentum
$\ell$.

We started from the observation that in the majority of the
physical applications of Schr\"{o}dinger equation in $D$
dimensions the relevant values of $\ell $ are usually small. In
this context, the DDT model itself is exceptional. In the strongly
spiked $\alpha \gg 1$ regime, it is {formally consistent} if and
only if the angular momenta $\ell$ are very high. In this regime,
the weak non-Hermiticity (supporting the real spectrum) {\em
requires} the presence of a strong centrifugal repulsive core. We
re-interpreted such a descriptive property of the model as its
internal, formal feature. Its {\em mathematical} consistency is
enhanced by the admissibility of the {complex} shifts and of a
${\cal PT}$ symmetric deformation of the axis of coordinates.

In this framework, our main purpose was to find a suitable
technique which would give the approximate low lying DDT spectrum
non-numerically. This effort was inspired by the enormous success
of the so called $1/\ell$ expansions, techniques which proved
extremely successful within the standard, Hermitian quantum
mechanics. Our study revealed that the transition between the
Hermitian and non-Hermitian models is entirely smooth. We
discovered, in particular, that the angular momentum parameter
$|\ell| \gg 1$ may serve as a guide to the introduction of the
suitable harmonic oscillator approximation of the low-lying (in
our case, DDT) spectrum.

We may conclude that the feasibility of the harmonic oscillator
approximation (presumably, not only in our non-Hermitian model
(\ref{rad})) is encouraging. We may expect that in the future the
more consequent and precise solution of the similar complex models
will prove obtainable by perturbative techniques. The
leading-order harmonic-oscillator construction will be followed by
the systematically constructed series of corrections in a way
which would parallel the $1/\ell$ expansions for the purely real
potentials.

\section*{Acknowledgments}

M. Z. supported by GA AS (Czech Republic), grant Nr. A 104 8004.
F. G. acknowledges support given by the Grant Agency of the Czech
Republic (grant No. 203/00/1025).

\section*{Figure captions}

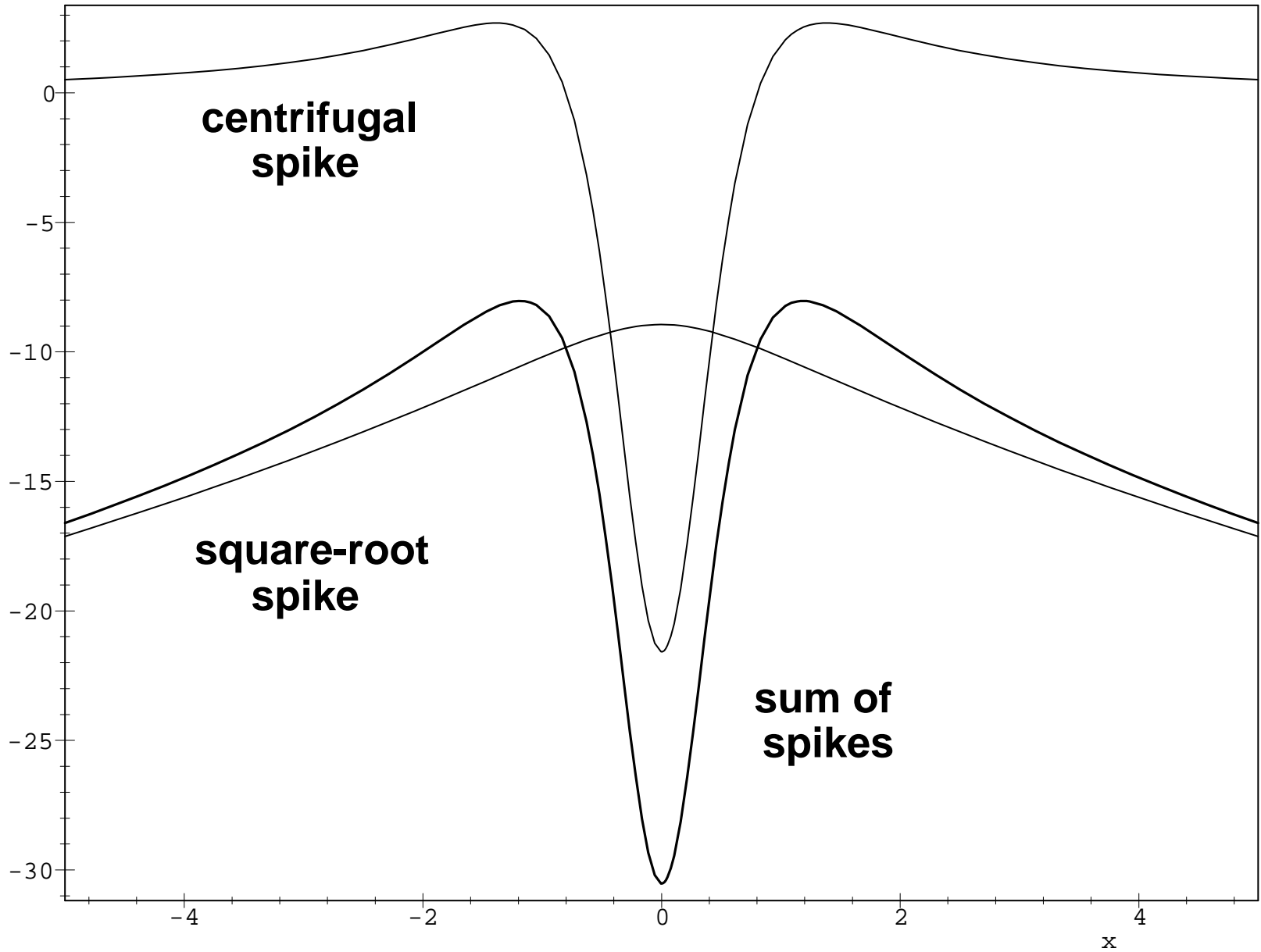
\noindent Figure 1.  Real parts of the spikes in eq. (\ref{rad})
at $\varepsilon = 0.8$.

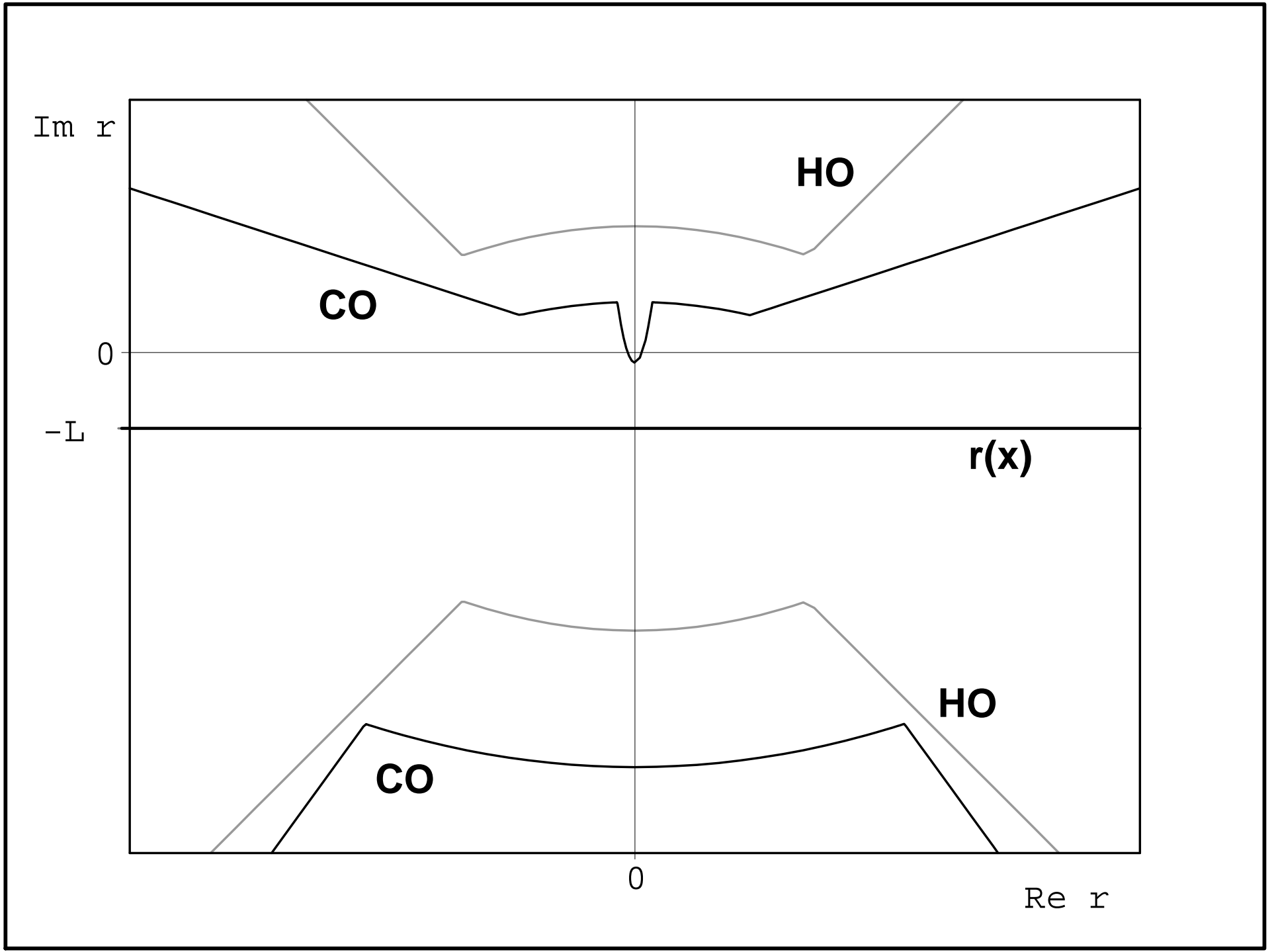
\noindent Figure 2. Optimal path ${r}(x)$ and asymptotic wedges:
boundary CO for the cubic oscillator (\ref{rad}) and boundary HO
for its harmonic-oscillator approximant (\ref{appr}).

\section*{Table captions}

\noindent Table 1. Energy levels of the solvable model
(\ref{adkv}) with $\ell = 5$.

\noindent Table 2. Energy levels of the generalized cubic model
(\ref{rad}) with $\alpha=0$ and growing $\ell$.

\noindent Table 3. The lowest three energy levels of the
generalized cubic model (\ref{rad}) with $\ell=10$ and various
$\alpha$.

\newpage

\begin{table}[p]
%===============

\caption{Energy levels of the solvable model (\ref{adkv}) with
$\ell = 5$.}

\begin{center}
\def\s{\hphantom{0}}
\begin{tabular}{cccc}
  \hline\hline
  exact    & numerical & large-$\ell$  & difference \\
  solution & solution  & approximation \\
  \hline
   -9 &  -9.00000 &  -8.954 & 0.046 \\
   -5 &  -5.00000 &  -4.954 & 0.046 \\
   -1 &  -1.00000 &  -1.954 & 0.046 \\
  \s3 & \s3.00000 & \s3.046 & 0.046 \\
  \s7 & \s7.00000 & \s7.046 & 0.046 \\
   11 &  11.00000 &  11.046 & 0.046 \\
   13 &  13.00000 &     --- &   --- \\
   15 &  15.00000 &  15.046 & 0.046 \\
   17 &  17.00000 &     --- &   --- \\
   19 &  19.00000 &  19.046 & 0.046 \\
  \hline\hline
\end{tabular}
\end{center}
\end{table}

\begin{table}[p]
%===============

\caption{Energy levels of the generalized cubic model (\ref{rad})
with $\alpha=0$ and growing $\ell$.}

\begin{center}
\def\s{\hphantom{0}}
\begin{tabular}{lccc}
  \hline\hline
  parameters & numerical & large-$\ell$  & difference \\
             & solution  & approximation \\
  \hline
  $\ell$=5  &   -11.52191 &   -11.390 & 0.132 \\
  L=1.821   &  \s-4.56482 &  \s-4.000 & 0.565 \\
            & \s\s1.87017 & \s\s3.390 & 1.520 \\
  \hline
  $\ell$=10 &   -28.76552 &   -28.686 & 0.079 \\
  L=2.361   &   -20.59867 &   -20.271 & 0.328 \\
            &   -12.70640 &   -11.855 & 0.851 \\
            &  \s-5.11663 &  \s-3.439 & 1.677 \\
            & \s\s2.14032 & \s\s4.976 & 2.836 \\
  \hline
  $\ell$=20 &   -68.72646 &   -68.680 & 0.046 \\
  L=3.086   &   -59.24706 &   -59.058 & 0.189 \\
            &   -49.91773 &   -49.435 & 0.482 \\
            &   -40.74589 &   -39.813 & 0.933 \\
            &   -31.73951 &   -30.191 & 1.549 \\
            &   -22.90712 &   -20.569 & 2.338 \\
            &   -14.25769 &   -10.947 & 3.311 \\
            &  \s-5.80054 &  \s-1.324 & 4.476 \\
            & \s\s2.45491 & \s\s8.298 & 5.843 \\
  \hline
  $\ell$=50 &  -211.13555 &  -211.113 & 0.023 \\
  L=4.427   &  -199.68009 &  -199.589 & 0.091 \\
            &  -188.29459 &  -188.065 & 0.230 \\
            &  -176.98040 &  -176.541 & 0.440 \\
            &  -165.73889 &  -165.017 & 0.722 \\
            &  -154.57149 &  -153.493 & 1.079 \\
            &  -143.47967 &  -141.969 & 1.511 \\
            &  -132.46494 &  -130.444 & 2.020 \\
            &  -121.52886 &  -118.920 & 2.608 \\
  \hline\hline
\end{tabular}
\end{center}
\end{table}

\begin{table}[p]
%===============

\caption{The lowest three energy levels of the generalized cubic
model (\ref{rad}) with $\ell=10$ and various $\alpha$.}

\begin{center}
\def\s{\hphantom{0}}
\begin{tabular}{cccc}
  \hline\hline
  $\alpha$ & numerical & large-$\ell$  & difference \\
           & solution  & approximation \\
  \hline
  \s20\s    &   -58.62190 &   -58.535 & 0.087 \\
            &   -49.83626 &   -49.533 & 0.303 \\
            &   -41.29014 &   -40.531 & 0.759 \\
  \hline
  10        &   -43.85223 &   -43.768 & 0.084 \\
            &   -35.40717 &   -35.083 & 0.324 \\
            &   -27.23003 &   -26.398 & 0.832 \\
  \hline
  0         &   -28.76552 &   -28.686 & 0.079 \\
            &   -20.59867 &   -20.271 & 0.328 \\
            &   -12.70640 &   -11.855 & 0.851 \\
  \hline
  -10       &   -13.35529 &   -13.282 & 0.073 \\
            &  \s-5.40717 &  \s-5.089 & 0.318 \\
            & \s\s2.27617 & \s\s3.104 & 0.828 \\
  \hline
  -20       & \s\s2.38131 & \s\s2.447 & 0.066 \\
            &  \s10.16462 &  \s10.463 & 0.298 \\
            &  \s17.70339 &  \s18.478 & 0.775 \\
  \hline\hline
\end{tabular}
\end{center}
\end{table}

\end{document}